\documentclass[prd,aps,showpacs,nofootinbib,amsmath,amssymb,eqsecnum,preprint,tightenlines]{revtex4}

\usepackage{amssymb}
\usepackage{graphicx}
\usepackage{bm} 
\usepackage{epsfig}
\usepackage[latin1]{inputenc}
\usepackage{float}

\def\lsim{\mathrel{\rlap{\lower4pt\hbox{$\sim$}}
    \raise1pt\hbox{$<$}}}                
\def\gsim{\mathrel{\rlap{\lower4pt\hbox{$\sim$}}
    \raise1pt\hbox{$>$}}}                
    
\begin{document}

\title{A Parallel Algorithm for Solving the 3d Schr\"odinger Equation}

\author{Michael Strickland and David Yager-Elorriaga}

\affiliation{Department of Physics \\ Gettysburg College \\
Gettysburg, PA 17325-1486 USA\\ }


\begin{abstract}
{
We describe a parallel algorithm for solving the time-independent 3d Schr\"odinger 
equation using the finite difference time domain (FDTD) method.  We introduce an optimized
parallelization scheme that reduces communication overhead between computational nodes.
We demonstrate that the compute time, $t$, scales inversely with the number of computational
nodes as $t \propto (N_{\rm nodes})^{-0.95 \pm 0.04}$.  This makes it possible to 
solve the 3d Schr\"odinger equation on extremely large spatial lattices using
a small computing cluster.
In addition, we present a new method for precisely determining the energy eigenvalues and
wavefunctions of quantum states based on a symmetry constraint on the FDTD initial condition.
Finally, we discuss the usage of multi-resolution techniques in order to speed up convergence
on extremely large lattices.
}
\end{abstract}

\maketitle

\section{Introduction}

Solving the 3d Schr\"odinger equation given an arbitrary potential $V({\vec r})$ is of great practical 
use in modern quantum physics; however, there are only a handful of potentials for which analytic
solution is possible.  In addition, any potential that does not have a high degree of symmetry,
e.g. radial symmetry, requires solution in full 3d, making standard ``point-and-shoot'' methods
\cite{numrepc} for solving one-dimensional partial differential equations of little use.  In this paper we 
discuss a parallel algorithm for solving the 3d Schr\"odinger equation given an arbitrary potential 
$V({\vec r})$ using the finite difference time domain (FDTD) method.

The FDTD method has a long history of application to computational electromagnetics  \cite{Yee:1966,Taflove:1980,Taflove:1990,Taflove:2005}.
In the area of computational electromagnetics parallel versions of the algorithms have been developed and tested
\cite{Varadarajan:1994,Liu:1994,Tinniswodd:1996,Sypniwski:2000,Palaniappan:2000,Yu:2005,Yu:2007}.
In this paper, we discuss the application of parallelized FDTD to the 3d Schr\"odinger equation.
The standard FDTD method has been applied to the 3d Schr\"odinger equation by several authors in the
past~\cite{feagin,roy1,wadehra,sullivan1,sullivan2,roy2,Sudiarta:2007}.  Here we show how to
efficiently parallelize the algorithm.  We describe our parallel algorithm for
finding ground and excited state wavefunctions and observables such as energy eigenvalues, and root-mean-squared radii.  Additionally, we introduce a way to use symmetry constraints for determining excited state wavefunctions/energies and 
introduce a multi-resolution technique that dramatically decreases compute 
time on large lattices. This paper is accompanied by an open-source release of a code that 
implements the algorithm detailed in this paper.  The code uses the Message Passing Interface (MPI)
protocol for message passing between computational nodes.  

We note that another popular method for numerical solution of the 3d Schr\"odinger equation is
the Diffusion Monte Carlo (DMC) technique, see \cite{Grimm:1971,Anderson:1975,Lester:1990,%
Mitas:1996,Foulkes:2001} and references therein.  The starting point for this method is the same as the FDTD
method applied here, namely transformation of the Schr\"odinger equation to imaginary time.  However, in the
DMC algorithm the resulting ``dynamical'' equations are transformed into an integral Green's function form 
and then the resulting integral equation is computed using stochastic sampling.  The
method is highly inefficient unless importance sampling \cite{Metropolis:1953,Umrigar:1993} is used.
DMC is efficiently parallelized and there are several codes which implement
parallelized DMC \cite{QMCBeaver:2004,QUMAX:2004,Aspuru-Guzik:2005}. 
The method is similar in many ways to the 
one presented herein; however, the method we use does not suffer from the fermion sign 
problem which forces DMC to use the so-called ``fixed-node approximation'' \cite{Anderson:1975}.
In addition, although the DMC algorithm can, in principle, be applied to extract properties of
the excited states of the system most applications to date only calculate the ground state wavefunction
and its associated expectation values.  The FDTD method described herein can extract both ground and excited
state wavefunctions.

The organization of the paper is as follows.  In Secs.~\ref{sec:setup} and \ref{sec:fdtdmethod} 
we briefly review the basics of the FDTD method applied to the 3d Schr\"odinger equation and
derive the equations necessary to evolve the quantum-mechanical wavefunction.  In 
Sec.~\ref{sec:initsymmetry} we discuss the possibility of imposing a symmetry constraint on
the FDTD initial condition in order to pick out different quantum-mechanical states.  In
Sec.~\ref{sec:strategy} we describe our strategy for parallelizing the FDTD evolution equations
and the measurement of observables.  In Sec.~\ref{sec:multiresolution} we introduce an 
efficient method of using lower-resolution FDTD wavefunctions as initial conditions for
higher-resolution FDTD runs that greatly speeds up determination of high-accuracy wavefunctions
and their associated observables.  In Sec.~\ref{sec:results} we give results for a few potentials
including benchmarks showing how the code scales as the number of computational nodes
is increased.  Finally, in Sec.~\ref{sec:conclusions} we conclude and give an outlook for
future work.

\section{Setup and Theory}
\label{sec:setup}

In this section we introduce the theory necessary to understand the FDTD 
approach for solving the time-independent Schr\"odinger equation.  
Here we will briefly review the basic idea of the FDTD method
and in the next section we will describe how to obtain the discretized ``equations of motion''.

We are interested in solving the time-independent Schr\"odinger equation with a static potential 
$V(\vec{r},t)=V(\vec{r})$ and a particle of mass $m$
\begin{equation}
E_n\psi_n(\vec{r})=\hat{H}\psi_n(\vec{r}) \, ,
\end{equation}
where $\psi_n$ is a quantum-mechanical wavefunction that solves this equation, 
$E_n$ is the energy eigenvalue corresponding to $\psi_n$, and
$\hat{H} = -\hbar^2\nabla^2/2m+V(\vec{r})$ is the Hamiltonian operator.  
In order to solve this time-independent (static) problem it is efficacious to consider
the time-dependent Schr\"odinger equation 
\begin{equation}
\label{schrod}
i\hbar\frac{\partial}{\partial t}\Psi(\vec{r},t) = \hat{H}\Psi(\vec{r},t) = 
\left[-\frac{\hbar^{2}}{2m}\nabla^{2}+V(\vec{r})\right]\Psi(\vec{r},t) \, .
\end{equation}
A solution to (\ref{schrod}) can be expanded in terms of the basis functions of the 
time-independent problem, i.e.
\begin{equation}
\Psi(\vec{r},t)= \sum_{n=0}^\infty {a_{n}\psi_{n}(\vec{r})e^{-iE_{n}t}} \, ,
\end{equation}
where $\{a_{n}\}$ are expansion coefficients which are fixed by initial conditions 
($n=0$ represents the ground state, $n=1$ the first excited state, etc.) and $E_n$ 
is the energy associated with each state.\footnote{The index $n$ is understood to 
represent the full set of quantum numbers of a given state of energy $E_n$.  In the 
degenerate case $\psi_n$ is an admixture of the different degenerate states.}

By performing a Wick rotation to imaginary time, $\tau = i t$, and setting $\hbar=1$ 
and $m=1$ in order to simplify the notation, we can rewrite Eq.~(\ref{schrod}) as
\begin{equation}
\label{wickeq}
\frac{\partial}{\partial \tau}\Psi(\vec{r},\tau)=\frac{1}{2}\nabla^{2}\Psi(\vec{r},\tau)
-V(\vec{r})\Psi(\vec{r},\tau) \, , 
\end{equation}
which has a general solution of the form
\begin{equation}
\Psi(\vec{r},\tau)= \sum_{n=0}^{\infty} {a_{n}\psi_{n}(\vec{r})e^{-E_{n}\tau}} \, .
\label{imagdecom}
\end{equation}

Since $E_{0}<E_{1}<E_{2}<...$, for large imaginary time $\tau$ the wavefunction 
$\Psi(\vec{r},\tau)$ will be dominated by the ground state wavefunction 
$a_{0}\psi_{0}(\vec{r})e^{-E_{0}\tau}$.  In the limit $\tau$ goes to infinity we have
\begin{equation}
\lim_{\tau \rightarrow \infty} \Psi(\vec{r},\tau)\approx a_{0}\psi_{0}(\vec{r})e^{-E_{0}\tau} \, ,
\end{equation}
Therefore, if one evolves Eq.~(\ref{wickeq}) to large imaginary times one will obtain a 
good approximation to the ground state wavefunction.\footnote{In this context a large 
imaginary time is defined relative to the energy splitting between the ground state and 
the first excited state, e.g. $e^{(E_0 - E_1)\tau} \ll 1$; therefore, one must evolve to imaginary
times much larger than $1/(E_1 - E_0)$.}

This allows one to determine the ground state energy by numerically solving equation 
(\ref{wickeq}) for large imaginary time, and then use this wavefunction to find the 
energy expectation value $E_{0}$:
\begin{equation}
\label{energy0}
E_0 = \frac{\langle\psi_{0}|\hat{H}|\psi_{0}\rangle}{\langle\psi_{0}|\psi_{0}\rangle}=\frac{\int\!d^{3}x\,\psi_{0}^{*}\hat{H}\psi_{0}}{\int\!d^{3}x\,|\psi_{0}|^{2}} \, ,
\end{equation}
However, the method is not limited to extraction of only the ground state wavefunction
and expectation values.  In the next sections we will describe two different methods that
can be used to extract, in addition, excited state wavefunctions.

\section{The Finite Difference Time Domain Method}
\label{sec:fdtdmethod}

To numerically solve the Wick-rotated Schr\"odinger equation (\ref{wickeq}) one can
approximate the derivatives by using discrete finite differences. For the application at
hand we can, without loss of generality, assume that the wavefunction is real-valued 
as long as the potential is real-valued. The imaginary time derivative becomes
\begin{equation}
\label{time}
\frac{\partial}{\partial \tau}\Psi(x,y,z,\tau)\approx\frac{\Psi(x,y,z,\tau+\Delta\tau)-\Psi(x,y,z,\tau)}{\Delta\tau}
\end{equation}
where $\Delta\tau$ is some finite change in imaginary time.

Similarly, the right hand side of equation (\ref{wickeq}) becomes
\begin{eqnarray}
\label{space}
\hspace{-2cm}\frac{1}{2}\nabla^{2}\Psi(\vec{r},\tau)-V(\vec{r})\Psi(\vec{r},\tau)&\approx&\nonumber\\
&&\hspace{-2.5cm}\frac{1}{2\Delta x^{2}}[\Psi(x+\Delta x,y,z,\tau)-2\Psi(x,y,z,\tau)+\Psi(x-\Delta x,y,z,\tau)]\nonumber\\
&&\hspace{-3cm}+\frac{1}{2\Delta y^{2}}[\Psi(x,y+\Delta y,z,\tau)-2\Psi(x,y,z,\tau)+\Psi(x,y-\Delta y,z,\tau)]\nonumber\\
&&\hspace{-3cm}+\frac{1}{2\Delta z^{2}}[\Psi(x,y,z+\Delta z,\tau)-2\Psi(x,y,z,\tau)+\Psi(x,y,z-\Delta z,\tau)]\nonumber\\
&&\hspace{-3cm}-\frac{1}{2}V(x,y,z)[\Psi(x,y,z,\tau)+\Psi(x,y,z,\tau+\Delta\tau)] \, ,
\end{eqnarray}
\\
where, in the last term, we have averaged the wavefunction in imaginary time
in order to improve the stability of the algorithm following Taflove \cite{Taflove:1990} and Sudiarta and Geldart \cite{Sudiarta:2008}.
Note that if the potential $V$ has singular points these have to be 
regulated in some way, e.g. by ensuring that none of the lattice points
coincides with a singular point.
Assuming, for simplicity, that the lattice spacing in each direction is
the same so that $a \equiv \Delta x = \Delta y = \Delta z$ this
equation can be rewritten
more compactly by defining a difference vector
\begin{eqnarray}
{\bf D} &\equiv& \frac{1}{a^2}[1,-2,1] \, ,
\end{eqnarray}
together with a matrix-valued ${\bf \hat\Psi}$ field
\begin{equation}
{\bf \hat\Psi} \equiv \left[ 
\begin{array}{ccc}
\Psi(x-a,y,z,\tau) & \Psi(x,y-a,z,\tau) & \Psi(x,y,z-a,\tau) \\
\Psi(x,y,z,\tau) & \Psi(x,y,z,\tau) &  \Psi(x,y,z,\tau)  \\
\Psi(x+a,y,z,\tau) & \Psi(x,y+a,z,\tau) & \Psi(x,y,z+a,\tau)
\end{array} \right] ,
\end{equation}
giving
\begin{eqnarray}
\label{spacesimple}
\hspace{-2cm}\frac{1}{2}\nabla^{2}\Psi(\vec{r},\tau)-V(\vec{r})\Psi(\vec{r},\tau)&\approx&\nonumber\\
&&\hspace{-2.5cm}\frac{1}{2}
\sum_{i=1}^3\;\left({\bf D\cdot\hat{\Psi}}\right)_i -\frac{1}{2}V(x,y,z)[\Psi(x,y,z,\tau)+\Psi(x,y,z,\tau+\Delta\tau)] \, ,
\end{eqnarray}

Rewriting equation (\ref{wickeq}) with equations (\ref{time}) and (\ref{spacesimple}) gives the following
update equation for $\Psi(x,y,z,\tau)$ in imaginary time:
\begin{equation}
\Psi(x,y,z,\tau+\Delta\tau) =  A \, \Psi(x,y,z,\tau) + \frac{B \, \Delta\tau}{2m} 
\, \sum_{i=1}^3\;\left({\bf D\cdot\hat{\Psi}}\right)_i \, ,
\label{updateeq}
\end{equation}
where $A$ and $B$ are
\begin{equation}
A\equiv\frac{1-\frac{\Delta\tau}{2}V(x,y,z)}{1+\frac{\Delta\tau}{2}V(x,y,z)} \, , \;\;\;\;\;
B\equiv\frac{1}{1+\frac{\Delta\tau}{2}V(x,y,z)} \, ,
\end{equation}
and we have reintroduced the mass, $m$, for generality.
Evolution begins by choosing a random 3d wavefunction as the initial condition.
In practice, we use Gaussian distributed random numbers with an amplitude of
one.  The boundary values of the wavefunction are set to zero; however, other
boundary conditions are easily implemented.~\footnote{We note that parallel 
implementations of absorbing boundary conditions may present a bottleneck
for the parallel calculation.  Many implementations of perfectly matched layers
exist \cite{Berenger:1994,Chew:1994,Gedney:1996}; however, only a few efficient parallel implementations exist
with the maximum efficiency of $t_{\rm compute} \propto N_{\rm nodes}^{-0.85}$ achieved
using the WE-PML scheme \cite{Zhou:2001,Rickhard:2002,Ramadan:2003,Ramadan:2007,Ramadan:2008}.
To the best of our knowledge the PML method has only been applied to unparallelized 
solution of the Schr\"odinger equation \cite{Fevens:1999,Alonso-Mallo:2003,Han:2004}.}  
Note that during the imaginary time evolution
the norm of the wavefunction decreases (see Eq.~\ref{imagdecom}), 
so we additionally renormalize the wavefunction during the evolution in order to avoid 
numerical underflow.  This does not affect physical observables.

We solve Eq.~(\ref{updateeq}) on a three-dimensional lattice with lattice spacing
$a$ and $N$ lattice sites in each direction.  Note that the lattice spacing $a$ and size 
$L = N a$ should be chosen so that the states one is trying to determine (i) fit inside
of the lattice volume, i.e. $\Psi_{\rm RMS} \ll L$, and (ii) are described with a sufficiently
fine resolution, i.e. $\Psi_{\rm RMS} \gg a$.  Also note that since we use an explicit method
for solving the resulting partial differential equation for the wavefunction,
the numerical evolution in imaginary time is subject to numerical instability if the time step is taken too large.  
Performing the standard von Neumann stability analysis \cite{Mitchell:1980} 
one finds that $\Delta \tau < a^2/3$ in order 
achieve stability.   For a fixed lattice volume $a = L/N$, therefore, $\Delta \tau \propto N^{-2}$ when 
keeping the lattice volume fixed.  The total compute time 
scales as $t_{\rm total} \propto N^3 N_{\rm time\; steps}$ and assuming $N_{\rm time\;steps} 
\propto (\Delta\tau)^{-1}$, we find that 
the total compute time scales as $t_{\rm total} \propto N^5$.

At any imaginary time $\tau$ the  energy of the state, $E$, can be computed 
via a discretized form of equation (\ref{energy0})
\begin{equation}
\hspace{16mm}
E[\Psi] = \frac{\sum_{x,y,z}\Psi(x,y,z,\tau) 
\Biggl[ \frac{1}{2} \sum_{i=1}^3\;\left({\bf D\cdot\hat{\Psi}}\right)_i -V(x,y,z)\Psi(x,y,z,\tau) \Biggr]}
{\sum_{x,y,z}\Psi(x,y,z,\tau)^{2}} \; .
\label{energysum}
\end{equation}

Excited states are extracted by saving the full 3d wavefunction to local memory
periodically, which we will call taking a ``snapshot'' of the wavefunction.  After 
convergence of the ground state wavefunction these snapshots can be used, one by 
one, to extract states with higher-energy eigenvalues by projecting out the ground state 
wavefunction, then the first excited state wavefunction, and so on~\cite{Sudiarta:2007}.  
In principle, one can extract as many states as the number of snapshots of the wavefunction 
saved during the evolution.  For example, assume that we have converged to the ground 
state $\psi_0$ and that we also have a snapshot version of the wavefunction 
$\Psi_{\rm snap}$ taken during the evolution.  To extract the first excited state $\psi_1$ 
we can project out the ground state using
\begin{equation}
|\psi_1\!> \,\; \simeq  \; |\Psi_{\rm snap}\!> -  \; |\psi_0\!><\!\psi_0|\Psi_{\rm snap}\!>  \, .
\end{equation}
For this operation to give a reliable approximation to $\psi_1$ the snapshot time should obey
$\tau_{\rm snap} \gg 1/(E_2 - E_1)$.
One can use another snapshot wavefunction that was saved
and obtain the second excited state by projecting out 
both the ground state and the first excited state.

Finally we mention that one can extract the binding energy of a state by computing
its energy and subtracting the value of the potential at infinity
\begin{equation}
E_{\rm binding}[\psi] = E[\psi] - \frac{<\!\psi|V_\infty|\psi\!>}{<\!\psi|\psi\!>} \, ,
\label{bindingenergy}
\end{equation}
where $V_\infty \equiv \lim_{r\rightarrow\infty} V(x,y,z)$ with $r = \sqrt{x^2+y^2+z^2}$ as usual.  
Note that if $V_\infty$ is a constant, then Eq.~(\ref{bindingenergy}) simplifies to $E_{\rm binding}[\psi] = 
E[\psi] - V_\infty$.

\section{Imposing symmetry conditions on the initial wavefunction}
\label{sec:initsymmetry}

Another way to calculate the energies of the excited states is to impose a symmetry
constraint on the initial conditions used for the FDTD evolution.  The standard
evolution calls for a random initial wavefunction; however, if we are solving a problem
that has a potential with sufficient symmetry we can impose
a symmetry condition on the wavefunction in order to pick out the different states
required.  For example, if we were considering a spherically symmetric Coulomb
potential then we could select only the $1s$, $2s$, $3s$, etc. states by requiring 
the initial condition to be reflection symmetric about the $x$, $y$, and $z$ 
axes.\footnote[1]{Of course, for a spherically symmetric potential a fully 3d method 
for solving the Schr\"odinger equation is
unecessary since one can reduce the problem to solving a 1d partial differential 
equation.  We only use this example because of its familiarity.}
This would preclude the algorithm finding any anti-symmetric states such as 
the $1p$ state since evolution under the Hamiltonian operator cannot break the 
symmetry of the wavefunction.  Likewise  to directly determine the
$1p$ excited state one can start by making the FDTD initial state wavefunction
anti-symmetric about one of the axes, e.g. the $z$-axis.  As we will show below this
provides for a fast and accurate method for determining the low-lying excited 
states.

Notationally, we will introduce two symbols, the symmetrization operator ${\cal S}_{i}$
and the anti-symmetrization operator ${\cal A}_{i}$.  Here $i$ labels the spatial
direction about which we are (anti-)symmetrizing, i.e. $i \in \{x,y,z\}$.  Although not
required, it is implicit that we perform the symmetrization about a plane with
$x=0$, $y=0$, or $z=0$, respectively.  In practice these are implemented by initializing
the lattice and then simply copying, or copying plus flipping the sign, elements from one
half of the lattice to the other.  In practice, we find that due to round-off error one should
reimpose the symmetry condition periodically in order to guarantee that lower-energy
eigenstates do not reappear during the evolution.

\section{FDTD Parallelization Strategy}
\label{sec:strategy}

\begin{figure*}[t]
\begin{center}
\includegraphics[width=15.2cm]{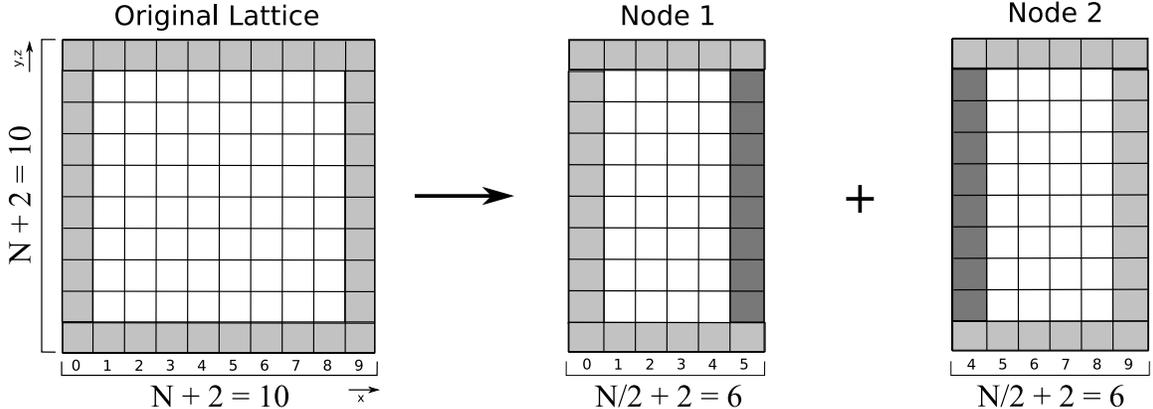}
\end{center}
\caption{
A sketch of the partition of an original lattice into two sub-lattices that can be simulated
on separate computational nodes.  In this example, we show the partitioning of an 
$N^3 = 8^3$ lattice into $M=2$ sub-lattices of $4\times8^2$.  The third dimension is
suppressed for clarity.  Light grey shaded boxes indicate
sites that contain boundary value information and the dark grey shaded boxes indicate
sites that contain information which must be synchronized between node 1 and node 2.
White boxes indicate regions where the updated wavefunction value is stored.
}
\label{fig:parallelizationsketch}
\end{figure*}

Parallelizing the FDTD algorithm described above is relatively straightforward.  Ideally,
one would segment the volume into $M$ equal subvolumes and distribute them equally 
across all computational nodes; however, in this paper we will assume a somewhat
simpler possibility 
of dividing the lattice into ``slices''. Our method here will be to start with a $N^3$ lattice and slice it 
along one direction in space, e.g. the $x$ direction, into $M$ pieces where $N$ is 
divisible by $M$.  We then
send each slice of $(N/M) \times N^2$ lattice to a separate computational node and
have each computational node communicate boundary information between nodes
which are evolving the sub-lattices to its right and/or left.  The partitioning of the lattice 
is indicated via a 2d sketch
in Fig.~\ref{fig:parallelizationsketch}.  In practice, in order to implement boundary
conditions and synchronization of boundaries between computation nodes compactly in the code, 
we add ``padding elements'' to the overall
lattice so that the actual lattice size is $(N+2)^3$.  The outside elements of the physical
lattice hold the boundary value for the wavefunction.  In all examples below the
boundary value of the wavefunction will be assumed to be zero; however,
different types of boundary conditions are easily accomodated.  When
slicing the lattice in order to distribute the job to multiple computational nodes we
keep padding elements on each slice so that the actual size of the slices is $(N/M + 2)
\times (N+2)^2$.  Padding elements on nodes that share a boundary are used to
keep them synchronized, while padding elements on nodes that are at the edges of the
lattice hold the wavefunction boundary condition.

\begin{figure*}[t]
\begin{center}
\includegraphics[width=13.2cm]{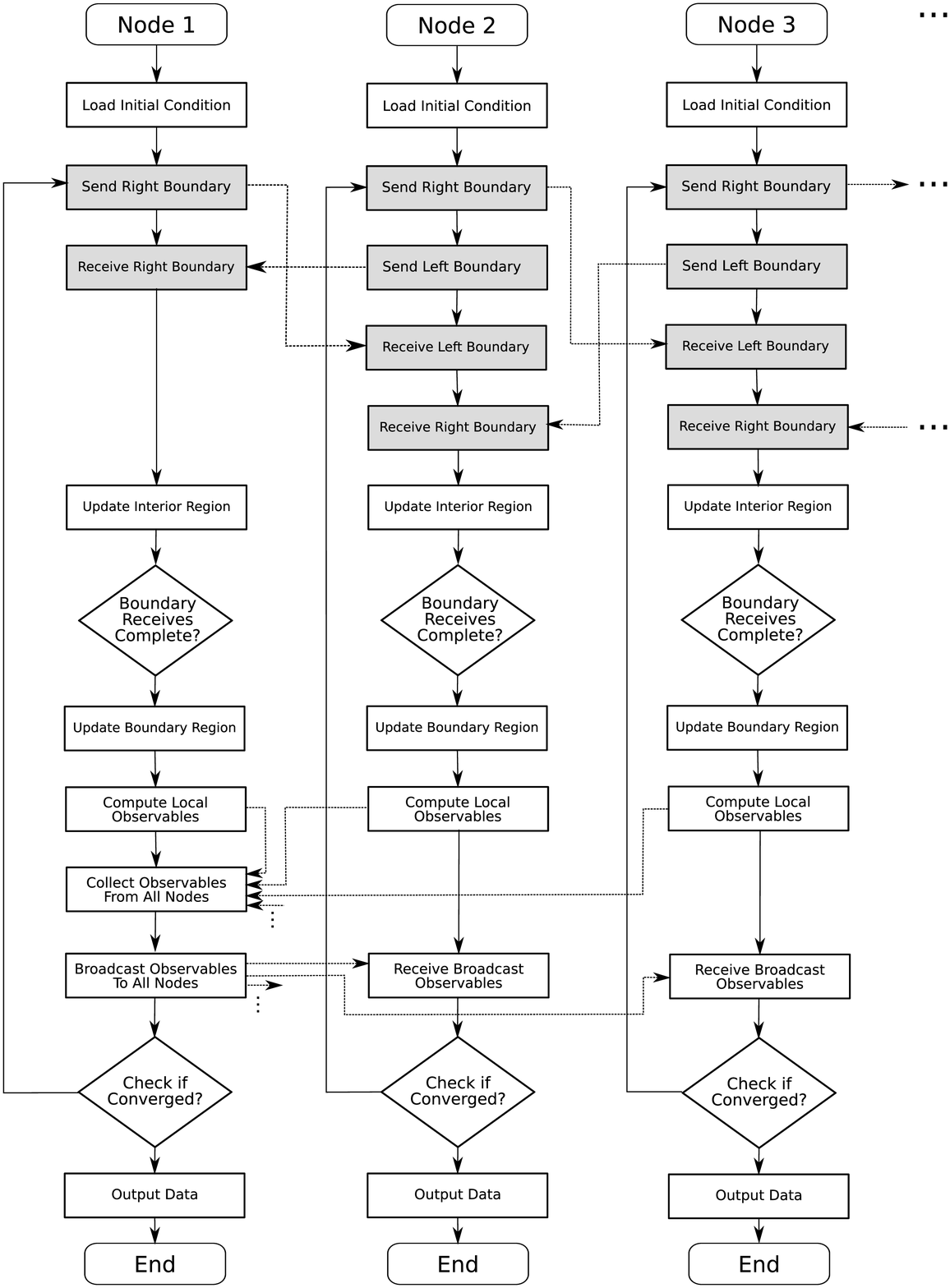}
\vspace{-6mm}
\end{center}
\caption{
Flow chart showing a sketch of our parallel algorithm.  Each column represents a 
distinct computational node.  Solid lines are process flow lines and dashed lines indicate 
data flow.  Shaded boxes indicate non-blocking communications calls that allow the 
process flow to continue while communications take place.}
\label{fig:programflow}
\end{figure*}

In Fig.~\ref{fig:programflow} we show a flow chart that outlines the basic method 
we use to evolve each node's sub-lattice in imaginary time.  In the figure each column
corresponds to a separate computational node.  Solid lines indicate the process flow
between tasks and dashed lines indicate data flow between computational nodes.
Shaded boxes indicate non-blocking communications calls that allow the process
flow to continue while communications take place.  As can be seen from
Fig.~\ref{fig:programflow} we have optimized each lattice update by making the first step 
in each update iteration a non-blocking send/receive between nodes.  While this 
send/receive is happening each node can then update the interior of its sub-lattice.  For 
example, in the two node case show in Fig.~\ref{fig:parallelizationsketch} this means that 
node 1 would update all sites with an $x$-index between 1 and 3 while node 2 
would update sites with $x$-index between 6 and 8.  Once these interior updates are
complete each node then waits for the boundary communication initiated previously
to complete, if it has not already done so.  Once the boundaries have been 
synchronized, the boundary elements themselves can be updated.  Going back to
our example shown in Fig.~\ref{fig:parallelizationsketch} this would mean that 
node 1 would update all sites with $x$-index of 4 and node 2 would update all
sites with an $x$-index of 5.

Convergence is determined by checking the ground state binding
energy periodically, e.g. every one hundred time steps, to see if it has changed
by more than a given tolerance.  In the code, the frequency of
this check is an adjustable parameter and should be tuned based on the expected 
energy of the state, e.g. if the energy is very close to zero then convergence can
proceed very slowly and the check frequency should be correspondingly larger.  
Parametrically the check frequency should scale as $1/E_0$.

For computation of observables each computational node computes its contribution
to the observable.  Then a parallel call is placed that collects the local values 
computed into a central value stored in computational node 1.  Then node 1 
broadcasts the value to the other nodes so that all nodes are then aware of the value
of the particular observable.  For example, to compute the energy of the state as
indicated in Eq.~(\ref{energysum}) each computational node computes the portion
of the sum corresponding to its sub-lattice and then these values are collected via
a parallel sum operation to node 1 and then broadcast out to each node.  Each node
can then use this information to determine if the wavefunction evolution is complete.  We
note that the normalization of the wavefunction is done in a similar way with each
node computing its piece of the norm, collecting the total norm to node 1, 
broadcasting the total norm to all nodes, and then each node normalizes the values
contained on its sub-lattice.  In this way computation of observables and wavefunction
normalization is also parallelized in our approach.

\subsection{Scaling of our 1d partitioning}
\label{sec:1dvs3d}

In most settings computational clusters are limited by their communication speed rather
than by CPU speed.  In order to understand how things scale we introduce two time scales:
$\Delta\tau_{u}$ which is the amount of time needed to update one lattice site and
$\Delta\tau_{c}$ which is the amount of time needed to communicate (send and receive) the information
contained on one lattice site.  Typically $\Delta\tau_{u} < \Delta\tau_{c}$ unless the cluster being used
is on an extremely fast network.  Therefore, the algorithm should be optimized to reduce the
amount of communications required.  

For the one-dimensional partitions employed here
\begin{eqnarray}
\tau_u &=& \left(\frac{N}{N_{\rm nodes}^{\rm 1d}}\right) N^2 \Delta\tau_u \, , \nonumber \\
\tau_c &=& 2 N^2 \Delta\tau_c  \, ,
\label{1dscaling}
\end{eqnarray}
where $N_{\rm nodes}^{\rm 1d}$ is the number of 1d slices distibuted across the cluster and
the factor of 2 comes from the 2 surfaces which must be communicated by the internal partitions.
For the calculation not to have a communications bottleneck we should have $\tau_c < \tau_u$.
Using (\ref{1dscaling}) we find that this constraint requires
\begin{equation}
N_{\rm nodes}^{\rm 1d} < \frac{1}{2} \left(\frac{\Delta\tau_{u}}{\Delta\tau_{c}}\right) N \, .
\end{equation}

In the benchmarks section below we will present measurements of $\Delta\tau_u$ and $\Delta\tau_c$
using our test cluster.  We find that $\Delta\tau_u \sim \Delta\tau_c/5$.  Using this, and assuming, as a
concrete example, a lattice size of $N=1024$ we find $N_{\rm nodes}^{\rm 1d} \lsim 102$.
For clusters with more than 102 nodes it would be more efficient to perform a fully 3d partitioning.
In the case of a fully 3d partitioning one finds that the limit due to communications overhead
is $N_{\rm nodes}^{\rm 3d} \lsim 39768$.

\section{The Multi-Resolution Technique}
\label{sec:multiresolution}

If one is interested in high-precision wavefunctions for low-lying excited states, 
an efficient way to do this is to use a multi-resolution technique.  This simply
means that we start with a random wavefunction on small lattice, e.g. $64^3$,
and use the FDTD technique to determine the ground state and first few 
excited states and save the wavefunctions, either in local memory or disk.  We
can then use a linear combination of the coarse versions of each state 
as the initial condition on a larger lattice, e.g. $128^3$, while keeping 
the lattice volume  fixed.  We can then ``bootstrap'' our way up to extremely large 
lattices, e.g. on the order of $1024^3 \rightarrow 2048^3$, by proceeding from 
low resolution to high resolution.  In the results section we will present 
quantitative measurements of the speed improvement that is realized using this 
technique.

\section{Results}
\label{sec:results}

In this section we present results obtained for various 3d potentials and
benchmarks that show how the code scales with the number of computational
nodes. Our benchmarks were performed on a small cluster of 4 servers, each with two 
quad-core 2 GHz AMD Opteron processors.  Each server can therefore efficiently 
run eight computational processes simultaneously, allowing a maximum of 32 
computational nodes. \footnote{Due to a server upgrade during publishing we were able
to extend to 64 computational nodes in the general benchmark section.}
The servers were networked with commercial 1 Gbit/s TCP/IP 
networking.  For the operating system we used 64 bit Ubuntu Server Edition 8.10 
Linux.

\subsection{Implementation}

In order to implement the parallel algorithm we use a mixture of C/C++ and 
the Message Passing Interface (MPI) library for message passing between 
computational nodes \cite{mpi}.  The servers used the OpenMPI implementation
of the MPI API.\footnote{The code was also tested against the MPICH implementation
of the MPI API with similar results.}  The code itself is open-sourced under the Gnu 
General Public License (GPL) and is available for internet download via the URL in 
Ref.~\cite{code}.

\subsection{General Benchmarks}

\begin{figure*}[t]
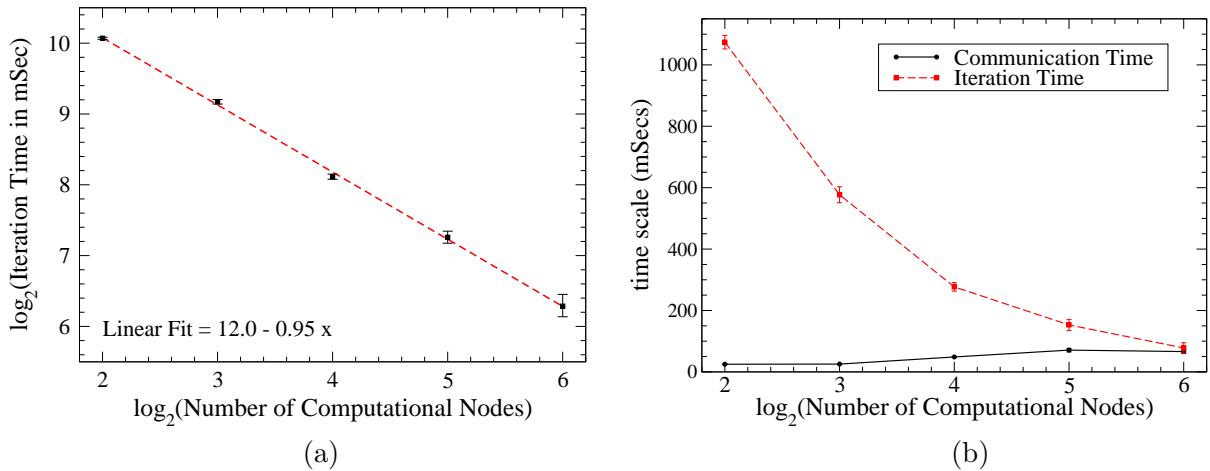

\vspace{1cm}
\begin{center}
\includegraphics[width=7.7cm]{itertimes.eps}$\;\;\;\;$
\includegraphics[width=7.7cm]{itercomm.eps}
\end{center}
\vspace{-5mm}
\hspace{1cm}(a)
\hspace{8cm}(b)
\caption{Time to complete one iteration of the update equation for the wavefunction
in imaginary time as a function of the number of nodes. In (a) the dashed line is 
a linear fit to the data.  In (b) the lines correspond to legend indicated.  Error bars are the standard error
determined by averaging observables over 10 runs, $\sigma_e = \sigma/\sqrt{N}$, where $\sigma$ is the
standard deviation across the sampled set of runs and $N$ is the number of runs.}
\label{fig:itertimes}
\end{figure*}

In this section we present data for the scaling of the time of one iteration and the
time for communication on a $N^3=512^3$ lattice.  As discussed in Sec.~\ref{sec:1dvs3d} 
we expect to see ideal scaling of the code as long as communication time is shorter than the 
update time, i.e. $\tau_c < \tau_u$.  In Fig.~\ref{fig:itertimes}a we show the time to
complete one iteration as a function of the number of computational nodes on a 
log-log axis along with a linear fit.  The linear fit obtained gives $\tau_{\rm iteration}
\propto N_{\rm nodes}^{-0.95\pm0.02}$.  In addition, in Fig.~\ref{fig:itertimes}b
we show a comparison of the full time for each iteration with the amount of time needed
to communicate a lattice site's information (in this case the local value of the wavefunction).
In both Fig.~\ref{fig:itertimes}a and \ref{fig:itertimes}b the error bars are the standard error
determined by averaging over 10 runs, $\sigma_e = \sigma/\sqrt{N}$, where $\sigma$ is the
standard deviation across the sampled set of runs and $N$ is the number of runs.

As can be seen from Fig.~\ref{fig:itertimes}b using a $512^3$ lattice the algorithm performs well up to
$N_{\rm nodes} = 64$ at which point the communication time becomes equal to the iteration
time.  For $N_{\rm nodes} > 64$ we would see a violation of the scaling above due to 
communication overhead.  Note that this is rough agreement with our estimate from 
Sec.~\ref{sec:1dvs3d} which, for $512^3$ lattice predicts the point where communications
and update times to be equal to be $N_{\rm nodes} \sim 51$.  Note that in  Fig.~\ref{fig:itertimes}b 
the increase in communication times as $N_{\rm nodes}$ increases is due to
the architecture of the cluster used for the benchmarks which has eight 
cores per server.  If $N_{\rm nodes} \leq 8$
then all jobs run on one server, thereby decreasing the communications overhead.  In the next section,
we will present benchmarks for different potentials in order to (a) confirm the scaling obtained above
in specific cases and (b) to verify that the code converges to the physically expected values for
cases which are analytically solvable.

\subsection{Coulomb Potential Benchmarks}
\label{sec:coulombbench}

We use the following potential for finding the Coulomb wavefunctions
\begin{equation}
V(r) = \left\{ \begin{array}{ll}
0 & \mbox{$\;\;\;\;r < a$} \\
-\frac{1}{r} + \frac{1}{a} & \mbox{$\;\;\;\;r\ge a$} \, ,
\end{array}
\right.
\label{coulombpot}
\end{equation}
where $a$ is the lattice spacing in units of the Bohr radius and $r$ is the
distance from the center of the 3d lattice.  The constant
of $1/a$ is added for $r \ge a$ in order to ensure
that the potential is continuous at $r=a$.  This is equivalent to making the potential
constant for $r \le a$ and shifting the
entire potential by a constant which does not affect the binding energy.  
Analytically, in our natural units
the binding energy of the $n$th state is $E_n = -1/(2(n+1)^2)$
where $n \ge 0$ is the principal quantum number labeling each state.  
The ground state therefore has a binding energy of $E_0 = -1/2$ and the
first excited state has $E_1 = -1/8$, etc.  Note that to convert these to electron
volts you should multiply by 27.2 eV.

\begin{figure*}[t]
\vspace{1cm}
\begin{center}
\includegraphics[width=10cm]{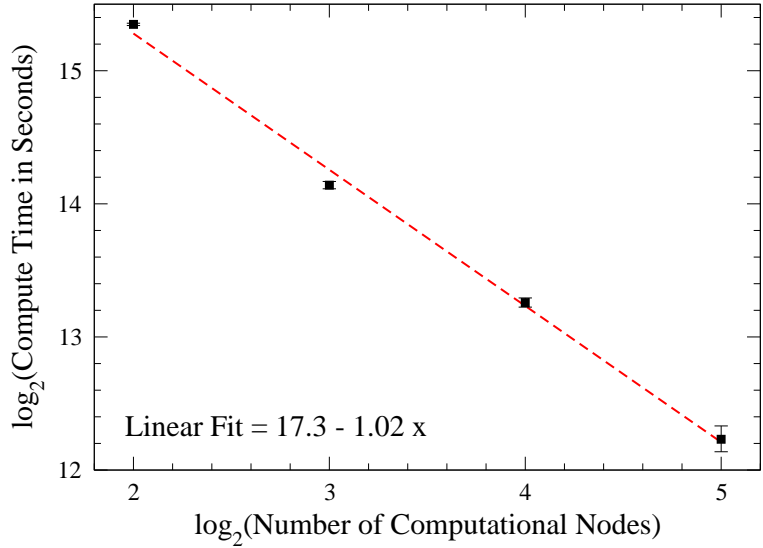}
\end{center}
\caption{Compute time versus number of computational nodes on a log-log plot for
the Coulomb potential specified in Eq.~(\ref{coulombpot}).  The dashed line is 
a linear fit to the data.  Scaling exponent indicates that, in this case, the compute time 
scales inversely with the number of compute nodes. Error bars are the standard error
determined by averaging over 10 runs, $\sigma_e = \sigma/\sqrt{N}$, where $\sigma$ is the
standard deviation across the sampled set of runs and $N$ is the number of runs.}
\label{fig:coulombscaling}
\end{figure*}

In Fig.~\ref{fig:coulombscaling} we show the amount of time needed in seconds to 
achieve convergence of the ground state binding energy to a part in $10^{6}$
as a function of the number of computational nodes for $N_{\rm nodes} \in \{4,8,16,32\}$
on a log-log plot.  For this benchmark we used a lattice with $N^3=512^3$, a constant lattice spacing 
of $a=0.05$, a constant imaginary time step of $\Delta\tau = a^2/4 = 6.25 \times 10^{-4}$,
and the particle mass was also set to $m=1$.
In order to remove run-by-run fluctuations due to the random initial conditions we used 
the same initial condition in all cases.
In Fig.~\ref{fig:coulombscaling} the error bars are the standard error
determined by averaging over 10 runs, $\sigma_e = \sigma/\sqrt{N}$, where $\sigma$ is the
standard deviation across the sampled set of runs and $N$ is the number of runs.
In all cases shown the first two energy levels obtained were $E_0 = -0.499$ and
$E_1 = -0.122$.  This corresponds to an accuracy of 0.2\% and 2.4\%, respectively.
In Fig.~\ref{fig:coulombscaling} the extracted scaling slope is close to 1 indicating that 
the compute time in this case scales almost ideally, i.e. inversely 
proportional to the number of computing nodes.  Note that the fit obtained in 
Fig.~\ref{fig:coulombscaling} has a slope with magnitude greater than 1 indicating scaling which is
better than ideal; however, as one can see from the figure there is some uncertainty
associated with this fit.

\subsection{3d Harmonic Oscillator Benchmarks}
\label{sec:hobench}

We use the following potential for finding the 3d harmonic oscillator wavefunctions
\begin{equation}
V(r) = \frac{1}{2} r^2 \, ,
\label{hopot}
\end{equation}
where $r$ is the distance from the center of the 3d lattice.

\begin{figure*}[t]
\vspace{1cm}
\begin{center}
\includegraphics[width=10cm]{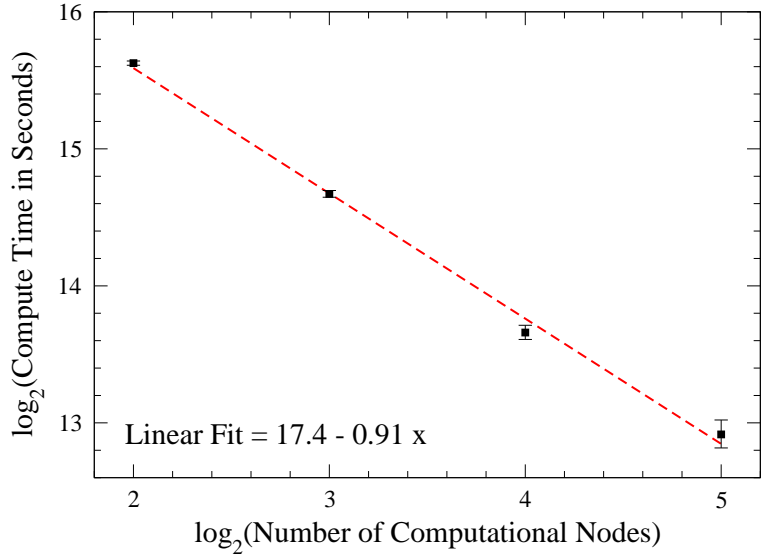}
\end{center}
\caption{
Compute time versus number of computational nodes on a log-log plot for
the 3d harmonic oscillator potential specified in Eq.~(\ref{hopot}).  The dashed line 
is a linear fit to the data.  Scaling exponent indicates that, in this case, the compute time 
scales as $t_{\rm compute} \propto N_{\rm nodes}^{-0.91}$.  Error bars are the standard error
determined by averaging over 10 runs, $\sigma_e = \sigma/\sqrt{N}$, where $\sigma$ is the
standard deviation across the sampled set of runs and $N$ is the number of runs.}
\label{fig:hoscaling}
\end{figure*}

In Fig.~\ref{fig:hoscaling} we show the amount of time needed in seconds to 
achieve convergence of the ground state binding energy to a part in $10^{6}$
as a function of the number of computational nodes for $N_{\rm nodes} \in \{4,8,16,32\}$.
For this benchmark we used a constant lattice spacing 
of $a=0.02$, a constant imaginary time step of $\Delta\tau = a^2/4 = 1.0 \times 10^{-4}$,
and a $N^3=512^3$ dimension lattice so that the box dimension was $L \equiv a N = 10.24$.
In Fig.~\ref{fig:hoscaling} the error bars are the standard error
determined by averaging over 10 runs, $\sigma_e = \sigma/\sqrt{N}$, where $\sigma$ is the
standard deviation across the sampled set of runs and $N$ is the number of runs.
The particle mass was also set to $m=1$.
In order to remove run-by-run fluctuations due to the random initial conditions we used 
the same initial condition in all cases.
In all cases the ground state energy obtained was $E_0 = 1.49996$ corresponding 
to an accuracy of 0.0026\%.
In Fig.~\ref{fig:hoscaling} the extracted scaling slope is 0.91 
meaning that the compute time scales as $t_{\rm compute} 
\propto N_{\rm nodes}^{-0.91}$ in this case.  This is a slightly different slope than in the Coulomb
potential case.  This is due to fluctuations in compute time due to server load and
sporadic network delays.
The scaling coefficient reported in the conclusions will be the average of all scaling
coefficients extracted from the different potentials detailed in this paper.

\subsection{Dodecahedron Potential}
\label{sec:dodecahedronbench}

The previous two examples have spherical symmetry and hence it is not necessary to apply
a fully 3d Schr\"odinger equation solver to them.  We do so only in order to show 
scaling with computational nodes and percent error compared to analytically 
available solutions.  As a nontrivial example of the broad applicability of the FDTD
technique we apply it to a potential that is a constant negative value of $V=-100$ inside a 
surface defined by a regular dodecahedron with the following 20 vertices
\begin{eqnarray}
\left(\pm \frac{1}{\phi},\pm \frac{1}{\phi},\pm \frac{1}{\phi}\right),
\left(0,\pm \frac{1}{\phi^2},\pm 1\right),
\left(\pm \frac{1}{\phi^2},\pm 1,0\right),
\left(\pm \frac{1}{\phi^2},0,\pm 1\right) \; ,
\label{dodecahedronpotential}
\end{eqnarray}
where $\phi = (1+\sqrt{5})/2$ is the golden ratio. The value -1 is mapped to the point
$n=1$ and the value 1 is mapped to the point $n=N$ in all three dimensions.  As a result,
the containing sphere has a radius of $\sqrt{3}(N-1)/2\phi$.

\begin{figure*}[t]
\begin{center}
\includegraphics[width=7cm]{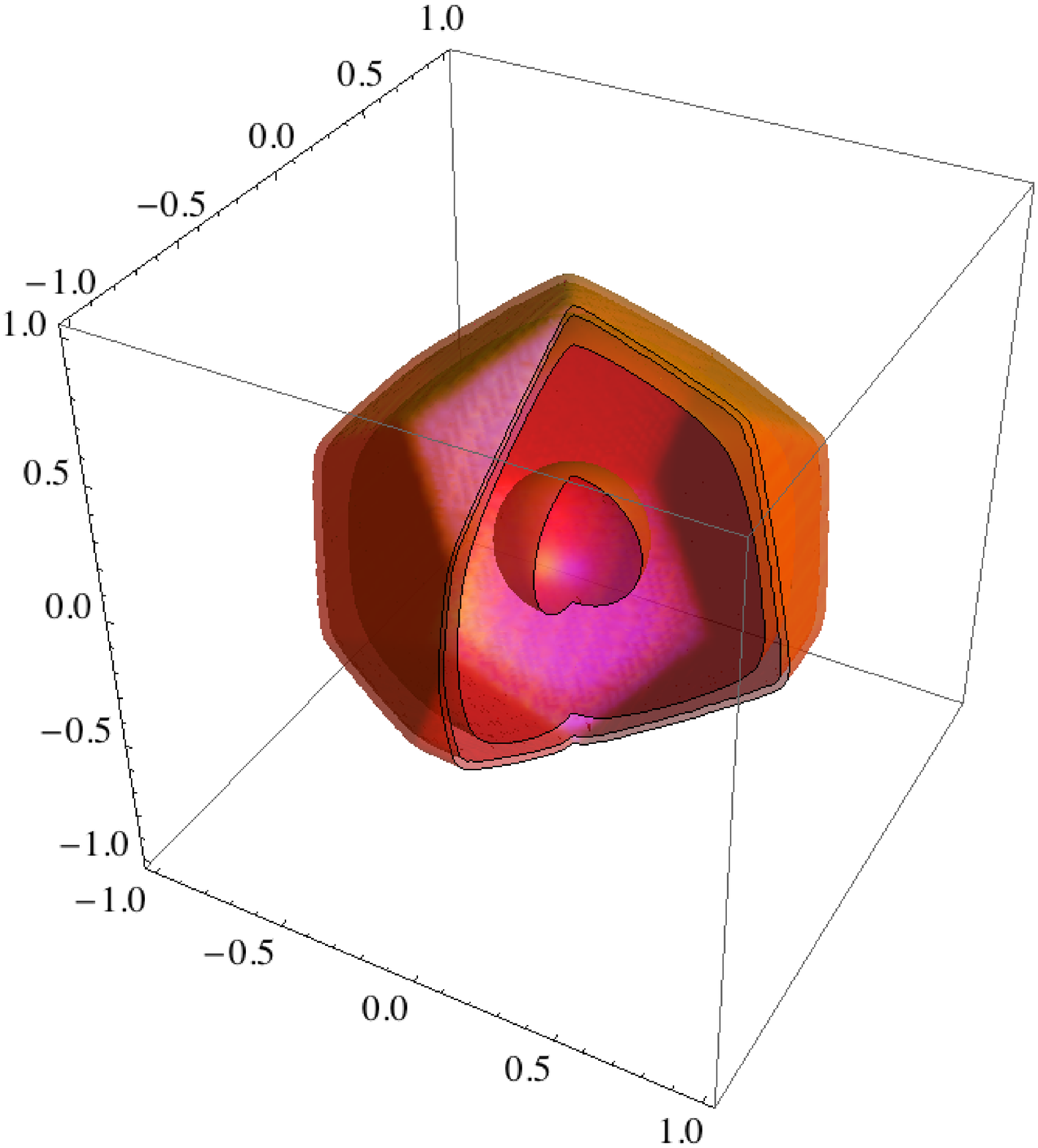}
$\;\;\;\;\;$
\includegraphics[width=7cm]{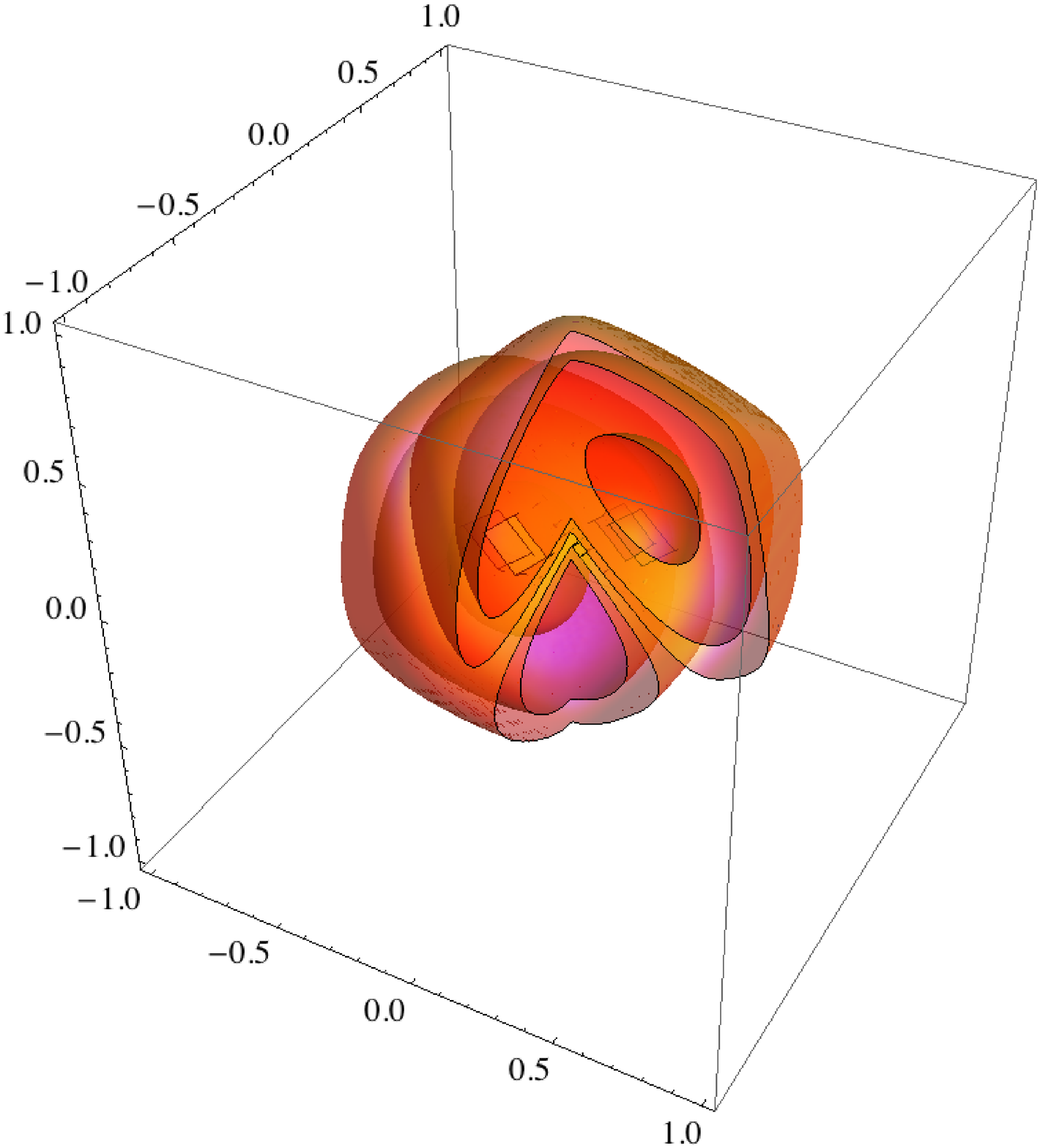}
\end{center}
\caption{
Ground state (left) and first excited state (right) of a dodecahedron potential.  Surfaces
are constant probability density surfaces.  For the ground state we show $\psi_0^2 \in \{10^{-11},
10^{-9},10^{-7},10^{-5}\}$ and for the first excited state $\psi_1^2 \in \{10^{-7},
10^{-6},10^{-5},10^{-4}\}$.  Positive quadrant defined by $x\le 0 \mid\mid y \ge 0$ is
cut out in order to view the interior of the wavefunctions.
}
\label{fig:dodecahedron}
\end{figure*}

In Fig.~\ref{fig:dodecahedron} we show the ground and first excited states
extracted from a run on a $128^3$ lattice with a lattice spacing of $a=0.1$,
an imaginary time step of $\Delta \tau = 0.001$ and particle mass of $m = 1$.
On the left we show the ground state and on the right the first excited state.
We find that the energies of these two levels are $E_0 = -99.78$ and $E_1=-99.55$.  
Note that for the first excited state the position of the node surface can change during
each run due to the random initial conditions used.  In practice, the node surface
seems to align along one randomly chosen edge of one of the pentagons that
make up the surface of the dodecahedron.

\begin{figure*}[t]
\vspace{1cm}
\begin{center}
\includegraphics[width=10cm]{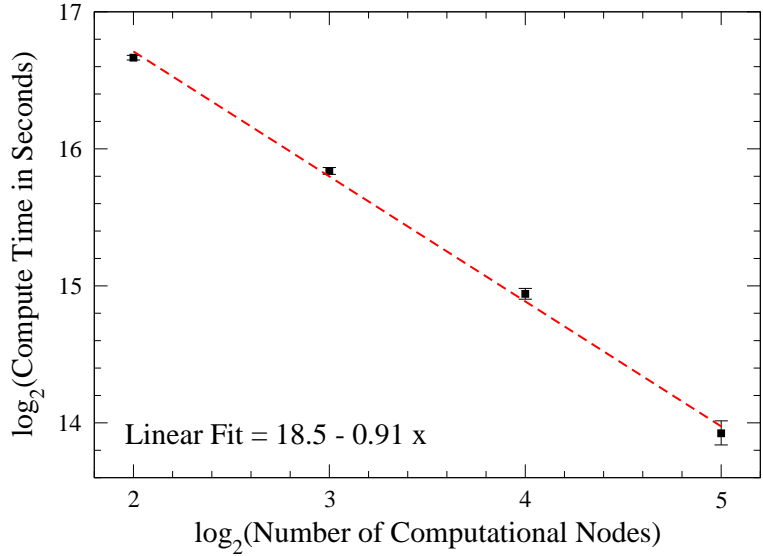}
\end{center}
\caption{
Compute time versus number of computational nodes on a log-log plot for
the dodecahedron potential using the vertices defined in Eq.~(\ref{dodecahedronpotential}).  
The dashed line is a linear fit to the data.  Scaling exponent indicates that, in this case, 
the compute time  scales as $t_{\rm compute} \propto N_{\rm nodes}^{-0.91}$.  Error bars are the standard error
determined by averaging over 10 runs, $\sigma_e = \sigma/\sqrt{N}$, where $\sigma$ is the
standard deviation across the sampled set of runs and $N$ is the number of runs.}
\label{fig:dodecahedronscaling}
\end{figure*}

In Fig.~\ref{fig:dodecahedronscaling} we show the amount of time needed in seconds to 
achieve convergence of the dodecahedron ground state binding energy to a part in $10^{6}$
as a function of the number of computational nodes for $N_{\rm nodes} \in \{4,8,16,32\}$.
For this benchmark we used a $512^3$ lattice with a constant lattice spacing of $a=0.1$, an 
imaginary time step of $\Delta \tau = 0.001$ and particle mass of $m = 1$.
In Fig.~\ref{fig:dodecahedronscaling} the error bars are the standard error
determined by averaging over 10 runs, $\sigma_e = \sigma/\sqrt{N}$, where $\sigma$ is the
standard deviation across the sampled set of runs and $N$ is the number of runs.
In all cases the ground state energy obtained was $E_0 = -99.97$.
In Fig.~\ref{fig:dodecahedronscaling} the extracted scaling slope is 0.91 
meaning that the compute time scales as $t_{\rm compute} 
\propto N_{\rm nodes}^{-0.91}$ in this case. 

\subsection{Applying Symmetry Constraints to the FDTD initial wavefunction}

One of the fundamental problems associated with using a single FDTD run to determine
both the ground state and excited states is that typically the excited states are much
more extended in space than the ground state, particularly for potentials with a ``long
range tail'' like the Coulomb potential.  For this reason it is usually difficult
to obtain accurate energy eigenvalues for both ground and excited states unless
the lattice has an extremely fine lattice spacing and a large number of points in
each direction so that the dimension of the box is also large.  In 
Sec.~\ref{sec:coulombbench} we presented benchmarks for the Coulomb potential
on a $512^3$ lattice that had a dimension of 25.6 Bohr radii.  As we found
in that section, we were able to determine the ground and first excited states to 0.2\% 
and 2.4\%.  Improving the accuracy of the first excited state would require going 
to a lattice with dimensions larger than $512^3$.  

While this is possible with the
parallelized code, there is a more efficient way to find excited states by applying
symmetry constraints to the initial wavefunction.  For example, to find the 1p state
of the Coulomb problem we can initialize the wavefunction as $\Psi_{\rm initial} = 
{\cal A}_z \Psi_{\rm random}$ as discussed in Sec.~\ref{sec:initsymmetry}.  
In this case we explicitly project out the ground
state wavefunction since it is symmetric about the $z$-axis.  Applying this method
on a $256^3$ lattice with lattice spacing $a=0.2$ and imaginary time step
$\Delta \tau = 0.01$ we find the first excited state energy to be $E_1 = -0.12507$ 
which is accurate to 0.06\%.  At the same time we can extract the next excited state 
which is anti-symmetric about the $z$-axis ($n=2$ state) finding in this case, 
$E_2 = -0.055360$, corresponding to an accuracy of 0.4\%.

The application of symmetry constraints can also allow one to pick out states
with different orientations in a 3d potential that breaks spherical symmetry.
In Ref~\cite{Dumitru:2009ni} this technique was used to accurately determine
the different heavy quarkonium p-wave states corresponding to angular momentum 
$L_z = 0$ and $L_z = \pm 1$.  Therefore, the ability to constrain the symmetry of 
the initial FDTD wavefunction is a powerful technique.

\subsection{Application of the Multi-resolution Technique}

In this section we present benchmarks for the application of the multi-resolution
technique to the Coulomb potential problem.  The current version of the code 
supports this feature by allowing users the option of saving the wavefunction
at the end of the run.  The saved wavefunctions can then be read in and used as
the initial condition for a subsequent run.  The saved wavefunctions can have
a different resolution than the resolution of the new run and the code automatically 
adjusts by sampling/spreading out the wavefunction appropriately.  

By using this technique we can accelerate the determination of the high accuracy
energy eigenvalues and wavefunctions.  In Sec.~\ref{sec:coulombbench} we found 
that using 32 computational nodes and a random initial wavefunction a $512^3$ run took 
approximately 1.3 hours.  Scaling naively to a $1024^3$ lattice, while keeping the lattice 
volume fixed, would take approximately 42 hours.  Using the multi-resolution 
technique and bootstrapping from $128^3$ up to $1024^3$ a high resolution 
ground state and energy eigenvalue can be computed in approximately 45 minutes 
using the same 
32 computational nodes.  At the final resolution of $a=0.025$ and a lattice size
of 25.6 Bohr radii the $1024^3$ run gives $E_0 = -0.499632$ which
is accurate to 0.07\%.  Therefore, the multi-resolution technique provides 
a performance increase of a factor of 50 compared to using random initial wavefunctions
for all runs.

\section{Conclusions and Outlook}
\label{sec:conclusions}

In this paper we have described a parallel FDTD algorithm for solving the 3d Schr\"odinger
equation.  We have shown that for large 3d lattices the method gives a 
compute time that scales as $t_{\rm compute} \propto N_{\rm nodes}^{-0.95 \pm 0.04}$.
This final scaling coefficient and associated error were obtained by averaging the three different
scaling coefficients extracted for the Coulomb, harmonic oscillator, and dodecahedron 
potentials.  
The  crucial optimization that allowed us to achieve nearly ideal scaling 
was the use of non-blocking sends/receives of the boundary data so that
update of each node's sub-lattice can proceed while communication of the boundary
information is taking place, providing for an ``inside-out'' update algorithm.

Additionally we introduced two novel techniques that can be used in conjunction with
the FDTD method.  First, we discussed the possibility of imposing a symmetry constraint
on the initial wavefunction used for the FDTD evolution.  The imposed symmetry
constraint allows us to easily construct states that are orthogonal to the ground state and/or
some other excited states.  Using this technique we can select states that have a certain
symmetry, thereby allowing for extremely accurate determination of the particular states we are
interested in.  Second, we introduced the ``multi-resolution technique'' which simply
means that we use the FDTD output wavefunctions from lower-resolution runs as the initial
condition for higher-resolution runs.  Using this method we showed that we can efficiently 
``bootstrap'' our way from small to large lattices, thereby obtaining high-accuracy wavefunctions
and eigenvalues in a fraction of the time required when using random initial wavefunctions on
large lattices.

The code developed for this paper has been released under an open-source GPL license \cite{code}.  
An obvious next step will be to extend the code to fully 3d partitions, which is in progress. 
Other areas of improvement include adding support for different types of boundary conditions 
and complex potentials \cite{forth}. 

\section*{Acknowledgements}
We thank V. Antocheviz Dexheimer, A. Dumitru, J. Groff, and J. Milingo for helpful comments.
M.S. thanks A. Dumitru, Y. Guo, and A. Mocsy for collaboration in the original 
project that prompted the development of this code.  M.S. also thanks Sharon Stephenson for 
support of D.Y. during the preparation of this manuscript.  D.Y. was supported by the 
National Science Foundation Award \# 0555652.

 

\end{document}